\documentclass[aps,prl,amsmath,amssymb,twocolumn,groupedaddress,
showpacs]{revtex4}

\usepackage{graphicx} 
\usepackage{dcolumn} 
\usepackage{bm}

\catcode`\° = \active 
\def°{\mbox{$^{\circ}$}}

\begin{document}

\title{Observation of a Long-Wavelength Hosing Modulation of
  a High-Intensity Laser Pulse in Underdense Plasma}

\author{M.~C.~Kaluza$^{1,2,3}$} 
\author{S.~P.~D.~Mangles$^1$}
\author{A.~G.~R.~Thomas$^{1,4}$} 
\author{Z.~Najmudin$^1$}
\author{A.~E.~Dangor$^1$} 
\author{C.~D.~Murphy$^{1,5,6}$}
\author{J.~L.~Collier$^5$} 
\author{E.~J.~Divall$^5$}
\author{P.~S.~Foster$^5$} 
\author{C.~J.~Hooker$^5$}
\author{A.~J.~Langley$^5$} 
\author{J.~Smith$^5$}
\author{K.~Krushelnick$^{1,4}$}

\affiliation{$^1$The Blackett Laboratory, Imperial College
  London SW7 2AZ, United Kingdom}

\affiliation{$^2$Institut f\"{u}r Optik und
  Quantenelektronik, Friedrich-Schiller-Universit\"{a}t,
  Max-Wien-Platz 1, 07743 Jena, Germany}

\affiliation{$^3$Helmholtz-Institut Jena, Helmholtzweg 4,
  07743 Jena, Germany}

\affiliation{$^4$Center for Ultrafast Optical Science
  (CUOS), University of Michigan, Ann Arbor, Michigan 48109,
  USA}

\affiliation{$^5$Central Laser Facility, Rutherford Appleton
  Laboratory, Oxon OX11 0QX, United Kingdom}

\affiliation{$^6$Clarendon Laboratory, University of Oxford,
  Oxford OX1 3PU, United Kingdom}


\begin{abstract}
  We report the first experimental observation of a
  long-wavelength hosing modulation of a high-intensity
  laser pulse.  Side-view images of the scattered optical
  radiation at the fundamental wavelength of the laser
  reveal a transverse oscillation of the laser pulse during
  its propagation through underdense plasma.  The wavelength
  of the oscillation $\lambda_{\rm hosing}$ depends on the
  background plasma density $n_{\rm e}$ and scales as
  $\lambda_{\rm hosing}\sim n_{\rm e}^{-3/2}$.  Comparisons
  with an analytical model and 2-dimensional
  particle-in-cell simulations reveal that this laser hosing
  can be induced by a spatio-temporal asymmetry of the
  intensity distribution in the laser focus which can be
  caused by a misalignment of the parabolic focussing mirror
  or of the diffraction gratings in the pulse compressor.
\end{abstract}

\maketitle

Since the first theoretical prediction \cite{pukhov02} and
experimental observation \cite{mangles04,geddes04,faure04}
of laser-accelerated electron pulses showing a narrow energy
spread, the prospect of a table-top laser-based particle
accelerator has come closer to reality. Such an electron
source would allow for the generation of THz-radiation
\cite{leemans03}, a high-intensity, narrowband femtosecond
source of visible or X-ray radiation \cite{schlenvoigt08,
  rousse04}, and possibly a free-electron laser in a
university-scale laboratory.  However, for all these
applications a high degree of stability and reproducibility
of the electron pulse is mandatory. This is difficult to
achieve due to the inherently non-linear processes involved
in the generation of quasi-monoenergetic electrons.  Here,
the ponderomotive potential of the high-intensity laser
pulse excites a plasma wave. Feedback between laser and
plasma wave induces longitudinal and transverse modulation
of the laser pulse and non-linear steepening of the wave.
When the wave eventually breaks plasma electrons are
injected into the electric field associated with the wave
(the ``wake field'') \cite{kaluza09} which can accelerate
them to relativistic energies.

To improve the stability of the electron beam, preionized
plasma channels \cite{leemans06, karsch07} or gas-filled
capillaries \cite{osterhoff08} have been used.
Additionally, improving the laser contrast \cite{mangles06}
and using two counter-propagation pulses \cite{faure06} has
increased the energy and pointing stability of the
electrons. However, asymmetries in the spatio-temporal
profile of the focussed laser pulse or of the initial plasma
distribution have been shown for pulses significantly longer
than the plasma wavelength to significantly affect the
propagation of the laser through the plasma and eventually
influence the interaction \cite{shvets94, sprangle94,
  duda99, najmudin03}.  Furthermore, the propagation of two
separated focal spots can be interlinked via the plasma
acting as a non-linear medium initiating a braided motion of
the two focal spots \cite{ren00}. Thus it is crucial to gain
as much information on the interaction process and about
influences of experimental parameters that can significantly
affect the interaction and eventually the electron-bunch
generation.

In this Letter, we show for the first time that also a laser
pulse with a length of the order of the plasma wavelength
exhibiting a distinctly asymmetric far-field intensity
distribution can trigger a hosing modulation leading to
transverse oscillations of the pulse during its propagation
through the plasma. The wavelength of this hosing strongly
depends on the background plasma density.

The experiments were carried out using the 10-TW Ti:Sapphire
laser system ASTRA at the Rutherford Appleton Laboratory.
45-fs, 350-mJ laser pulses at a central wavelength of
$\lambda_{\rm L}=800$\,nm were focussed into the leading
edge of a supersonic He-gas jet produced by a conical nozzle
having an exit diameter of 2\,mm. This gas jet exhibited an
almost flat-top longitudinal density profile as
characterized before the experiment. The density in the
plateau region could be adjusted by changing the backing
pressure.  Assuming full ionization the corresponding plasma
densities were between $n_{\rm e}=4\times10^{18}$ and
$4\times10^{19}$\,cm$^{-3}$.  The horizontally polarized
laser pulses were focussed by an $f/16$ off-axis parabolic
mirror to a focal spot with a diameter of 30\,$\mu$m, the
intensity averaged within this area was
$1.2\times10^{18}$\,W/cm$^2$. The fluence profile of the
focussed laser pulses was strongly attenuated and imaged
onto a CCD camera. During the full-power shots, the pointing
and an equivalent fluence distribution of the far-field of
the laser could be monitored by imaging the leakage of the
laser through the last plane mirror in front of the parabola
using an $f/20-$lens and a CCD. Looking in the direction of
laser polarization the interaction region in the plasma was
imaged using a high-quality $f/2-$lens with a magnification
of 10 onto a 12-bit CCD camera equipped with an interference
filter transmitting at $(800\pm10)$\,nm.  The spatial
resolution was close to 2\,$\mu$m, the field of view in the
interaction region was 540\,$\mu$m $\times$ 400\,$\mu$m.
MeV-electrons generated during the interaction could be
detected by an electron beam monitor consisting of a CCD
camera looking at a fluorescent screen positioned 480\,mm
behind the target and shielded by 20\,mm thick aluminium
making it sensitive to electrons with $E_{\rm
  kin}\ge11\,$MeV.  Furthermore, the spectra of the
electrons could be measured using a high-resolution magnetic
spectrometer equipped with imaging plate detectors.  The
electron spectra exhibited quasi-monoenergetic features.

\begin{figure}[h]
  \begin{center}
    \includegraphics[width=0.45\textwidth]{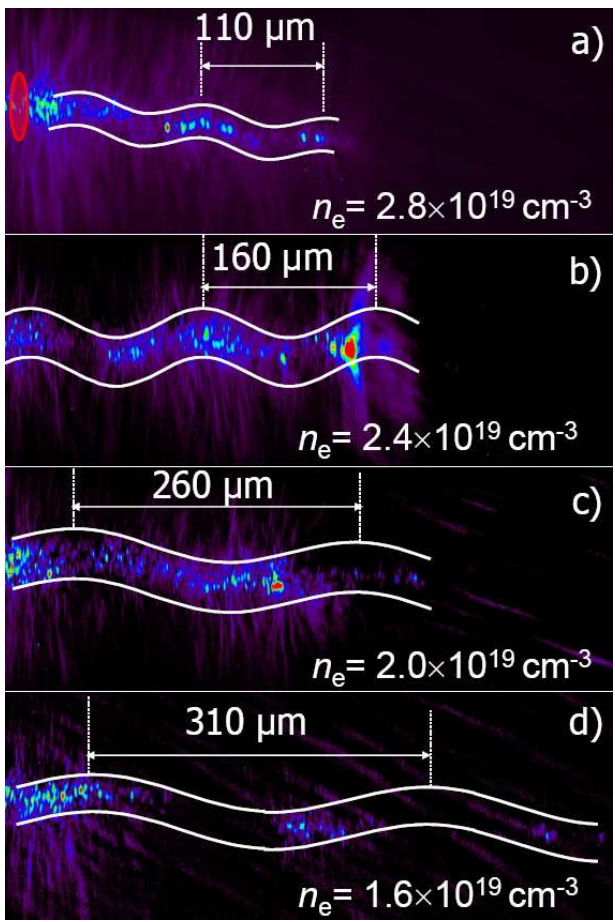}
    \caption{\label{fig:Hosing_experimental}
      Side-view images of the electromagnetic radiation
      scattered from the plasma at $\lambda_{\rm L}$. The
      images have a size of $500\,\mu$m$\times100\mu$m and
      have been stretched vertically by a factor of 2.  The
      laser pulse shows a hosing-type transverse
      oscillation, highlighted by the white lines. The
      hosing wavelength continuosly increases with
      decreasing plasma density $n_{\rm e}$.}
  \end{center} 
\end{figure}

Fig.~\ref{fig:Hosing_experimental} shows four images of the
scattered light recorded at the fundamental wavelength by
the side-view imaging system.  
All images show a field-of-view of
$500\,\mu$m$\times\,100\,\mu$m.  Note that they have been
stretched vertically by a factor of 2.  The initial plasma
densities in these four images were $n_{\rm e}=2.8$ (a),
$2.4$ (b), $2.0$ (c), and $1.6\times10^{19}$\,cm$^{-3}$ (d),
respectively. All images have been filtered differently to
record similar dynamic ranges for different levels of the
scattered light.

The laser pulse (propagating left to right) was focused into
the leading edge of the gas jet located close to the left
boundary of each image.  The red ellipse in
Fig.\,\ref{fig:Hosing_experimental}\,a) depicts the size and
approximate position of the vacuum laser focus. As the laser power is high enough to trigger relativistic
self-focussing (the critical power for a density of $n_{\rm
  e}=2.8\times10^{19}$\,cm$^{-3}$ is 1.2\,TW, well below our
laser power), the transverse diameter of the laser pulse
decreased during its propagation through the plasma.  Note
that since the radiation is only scattered from the plasma
electrons at the instantaneous position of the laser pulse
in the plasma, each image represents a history of the
instantaneous position of the laser as a function of
propagation distance. Hence, the horizontal axis represents
both distance and time of propagation.

\begin{figure}[h]
  \begin{center}
    \includegraphics[width=0.44\textwidth]{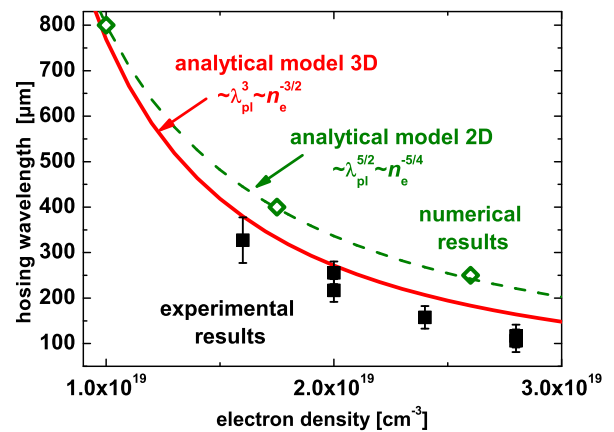}
    \caption{\label{fig:Hosing_wavelength}
      Dependence of the hosing wavelength $\lambda_{\rm
        hosing}$ on the electron density $n_{\rm e}$ for the
      experiment (black solid squares) and for the results
      from the numerical simulation (green open diamonds).
      These results are compared to predictions by an
      analytical model scaling like $\lambda_{\rm
        hosing}\sim n_{\rm e}^{-3/2}$ in 3D (red line) and
      like $\lambda_{\rm hosing}\sim n_{\rm e}^{-5/4}$ in 2D
      (green line).}
  \end{center} 
\end{figure}

Furthermore, it is obvious in all four images that the laser
pulse does not propagate on a straight line but carries out
transverse hosing-type oscillations. While the amplitude of
the oscillation of a few $\mu$m remains approximately
constant both during the propagation and for the different
background densities, the wavelength of this hosing
modulation significantly increases with decreasing plasma
density from $(110\pm25)\,\mu$m for the highest density up
to $(310\pm50)\,\mu$m for the lowest. This is summarized in
Fig.~\ref{fig:Hosing_wavelength}.
Note that due to the decreasing plasma density the intensity
of the scattered light decreases, since the scattering is
directly proportional to the number of free electrons. It
had decreased below our detection threshold for even lower
densities.

The long-wavelength hosing modulation observed in our
experiment is very likely to be caused by the focal spot of
the laser pulse exhibiting a spatio-temporal asymmetry in
its intensity distribution relative to the phase fronts of
the pulse. This asymmetry can arise due to a variety of
reasons in an chirped pulse amplification laser system, such
as by a slight misalignment of the parabolic focussing
mirror, a slight misalignment of the diffraction gratings
used to compress the pulse after amplification, pointing
fluctuations of the laser pulse itself induced by mechanical
vibrations or air turbulences or a combination of these
effects.
\begin{figure}[h]
  \begin{center}
    \includegraphics[width=0.48\textwidth]{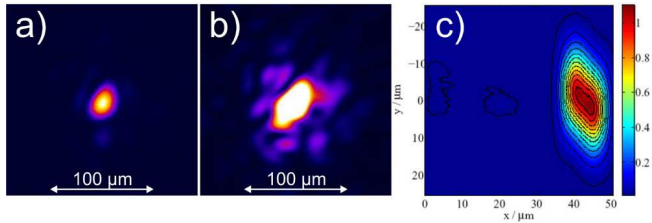}
    \caption{\label{fig:LaserPulseProfile}
      Energy-density profile of the laser far-field at two
      different filter levels, a) and b).  Snapshot of the
      initial, asymmetric laser-intensity profile used in
      the simulations, c).}
  \end{center} 
\end{figure}
Due to a feedback mechanism between the intensity
distribution of the laser pulse and the plasma density
mediated by the local refractive index of the plasma,
$\eta({\bf r})=(1-n_{\rm e}({\bf r})/n_{\rm cr})^{1/2}$, and
the ponderomotive potential of the laser, the propagation of
the laser pulse is strongly influenced by the plasma density
distribution. Here, $n_{\rm cr}=\omega_{\rm
  L}^2\varepsilon_0 m_{\rm e}/e^2=4\pi^2\varepsilon_0m_{\rm
  e}c^2/e^2\lambda_{\rm L}^2$ is the critical density of the
plasma for the wavelength $\lambda_{\rm L}$ and frequency
$\omega_{\rm L}$ of the laser light, $\varepsilon_0$ is the
dielectric constant and $e$ and $m_{\rm e}$ are the electron
charge and mass, respectively. The gradient of the plasma's
refractive index initiated by the head of the pulse via its
ponderomotive potential will be asymmetric with respect to
the bulk of the pulse, and will therefore provide an
effective focussing ``force''.  This exerts a transverse,
restoring force on the bulk of the pulse, causing this part
of the pulse to execute transverse oscillations. Typical,
slightly asymmetric laser foci measured during the
experiment with different filter levels are shown in
Fig.\,\ref{fig:LaserPulseProfile}\,a) and b).  Note that
both images show a time-integrated measurement of the
intensity distribution, i.e. the laser fluence in the focal
plane. Under conditions where no such focal spot asymmetry
was present, no hosing modulation could be detected.

\begin{figure}[h]
  \includegraphics[width=0.48\textwidth]{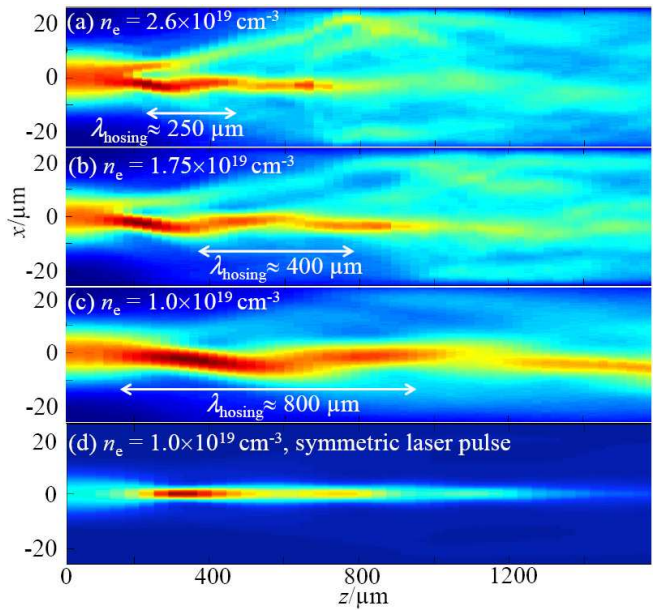}
  \begin{center}
    \caption{\label{fig:Hosing_numerical}
      Numerically simulated propagation of the laser pulse
      in the plasma. The images show the energy-density
      distribution the scattered light at the fundamental
      wavelength in transverse direction for electron plasma
      densities of $2.6\times10^{19}$\,cm$^{-3}$ (a),
      $1.75\times10^{19}$\,cm$^{-3}$ (b), and
      $1.0\times10^{19}$\,cm$^{-3}$ (c) and (d). The
      corresponding hosing wavelengths are also indicated.
      No hosing occurs for a symmetric laser pulse (d).}
  \end{center} 
\end{figure}

To verify this hypothesis, we carried out 2-dimensional
Particle-In-Cell (2D-PIC) simulations using the code {\sc
  Osiris} \cite{fonseca02} starting with an asymmetric laser
pulse. The spatio-temporal pulse intensity profile used in
our simulations, as shown in
Fig.~\ref{fig:LaserPulseProfile} c) was achieved by
overlapping two laser pulses of equal intensity, both having
a radius of 10\,$\mu$m ($1/e^2$ in intensity) mimicking the
focal-spot asymmetry in the experiment. The two parts of the
pulse were separated both in longitudinal (i.e.~in
propagation direction) and in transverse direction by
$6\lambda_{\rm L}$. They were overlapped at focus, so that
the phase fronts were flat. This resulted in a smooth pulse
profile despite being constructed of two pulses. The size of
each part of the focal spot in the simulation was chosen as
$w_0=10\,\mu$m, which is slightly smaller than in the
experiment to account for the under-estimated self-focussing
in 2D simulations.  Accordingly, $a_0=1.5$ was chosen higher
than in the experiment.
Fig.~\ref{fig:Hosing_numerical}\,(a)-(c) show the numerical
results obtained for three different densities.
It is obvious that the propagation of the laser pulse shows
a transverse hosing modulation, with amplitude and
wavelength similar to the experiment as also shown in
Fig.~\ref{fig:Hosing_wavelength}.
Fig.~\ref{fig:Hosing_numerical}\,(d) shows the result for
similar parameters as in (c) but for a pulse exhibiting no
spatio-temporal asymmetry.  In this case no hosing occurs.

A heuristic description of the interaction can be found by
assuming that the bulk of the pulse resides in the
quasi-static wake generated by the head of the pulse due to
its ponderomotive force. The general approach is to assume
the pulse is weakly relativistic, so that a density
depression but not complete cavitation is generated by the
pulse-front. The bulk of the pulse, displaced transversely
to the propagation direction with respect to the head of the
pulse, experiences the refractive index gradient associated
with the density depression. This acts to direct the pulse
energy towards the axis.

By considering the bending of phase fronts of the bulk of
the pulse analogous to \cite{mori97}, the position of the
centroid of the bulk of the pulse, $\langle r\rangle$, with
respect to the radially varying plasma density induced by
the front of the pulse, neglecting longitudinal dynamics,
can be expressed as
\begin{equation}
  \frac{\partial^2}{\partial t^2}\langle
  r\rangle\approx-\frac{c^2}{2}\frac{\partial }{\partial
    r}\left(\frac{\delta n_{\rm e}}{n_{\rm cr}}\right).
\end{equation}
Assuming an approximate balance between the ponderomotive
and wakefield potentials for a weakly relativistic pulse,
the density perturbation can be expressed as $\delta n_{\rm
  e}/n_{\rm cr}\approx(c^2/\omega_{\rm L}^2) \nabla^2
a^2/4$. If the field strength is distributed as
$a(r)^2=a_0^2(w_0^2/w^2)\cos(\pi r/2w)$, where $w_0$ and
$a_0=eE_0/(m_{\rm e}c\omega_{\rm L})$ are the initial
pulse-front radius and the normalized amplitude of the
vector potential, respectively ($E_0$ is the amplitude of
the laser's electric field), and $w$ is the radius of the
front of the pulse after self-focussing, then $\delta n_{\rm
  e}/n_{\rm cr}\approx-(a_0^2w_0^2\pi^2c^2/16\,\omega_{\rm
  L}^2w^4)\cos(\pi r/2w)$. The expression for the field
strength neglects absorption of pulse energy which conserves
laser power, $a_0w_0=aw$. We will assume that the pulse
asymmetry lies in the $x-z$ plane so that the oscillations
will occur in the $x$-direction.  Hence, for the bulk of the
pulse the equation for the centroid position is now
\begin{equation}
  \frac{\partial^2}{\partial t^2}\langle x\rangle\approx
  -\frac{a_0^2w_0^2\pi^3c^4}{64\omega_{\rm L}^2w^5}\sin\left(
    \frac{\pi}{2w}\langle x\rangle\right).
\end{equation}
To first order in $\langle x\rangle$, this is a harmonic
oscillator equation with a frequency $\omega_{\rm
  hosing}\approx{a_0w_0\pi^2c^2}/({8\sqrt{2}\omega_{\rm
    L}w^3})$.  Using this frequency of oscillation, the
expression yields $\lambda_{\rm
  hosing}\approx{16\sqrt{2}\omega_{\rm L}w^3}/({a_0w_0\pi
  c})$.  For a weakly relativistic pulse, the pulse
self-focusses to a radius of approximately half the plasma
wavelength, $w\approx\pi c/\omega_{\rm p}$ \cite{thomas07},
leading to
\begin{equation}
  \lambda_{\rm hosing}\approx
  \frac{4\sqrt{2}\lambda_{\rm L}^2}{a_0w_0}\left(\frac{n_{\rm
  cr}}
    {n_{\rm e}}\right)^{3/2}.
\end{equation}
For initial pulse radius and field strength $w_0=15\,\mu$m,
$a_0=0.75$, respectively, and $\lambda_{\rm L}=0.8\,\mu$m, a
hosing wavelength of $\lambda_{\rm hosing}=0.32\mu{\rm
  m}\cdot({n_{\rm cr}}/{n_{\rm e}})^{3/2}$ is obtained. This
agrees well with the experiment as shown by the red line in
Fig.~\ref{fig:Hosing_wavelength}. For the simulations, in a
2D slab geometry, the expression for constant power is
$a_0^2w_0=a^2w$. This modifies the hosing wavelength to be
$\lambda_{\rm hosing, 2D}\approx8\lambda_{\rm
  L}^{3/2}/(a_0\sqrt{w_0}) ({n_{\rm cr}}/{n_{\rm
    e}})^{5/4}$, which explains the slightly different
scaling. Using $w_0$=10\,$\mu$m, $a_0$=1.5 gives the dashed
green line in Fig.~\ref{fig:Hosing_wavelength}, which shows
excellent agreement between the model and the 2D-simulation
data.

In conclusion, we have observed the transverse hosing-type
oscillation of an ultra-short laser pulse during its
propagation through underdense plasma.  Analytical estimates
and 2D-PIC simulations have shown that such a transverse
oscillation can be caused by a spatio-temporal asymmetry of
the intensity of the laser pulse.  For experiments towards
the generation and application of collimated,
quasi-monoenergetic electron pulses where stability is an
important issue, such transverse oscillations of the laser
pulse are likely to influence the electron acceleration
process. To improve the stability of the electron pulse it
is necessary to carefully control and monitor the
spatio-temporal characteristic of the driving laser pulse.
It is conceivable that the hosing modulation will become
more pronounced at higher laser powers as in this case the
sensitivity to phase front tilts is likely to be higher.
Deliberately exciting hosing of this form could also have a
significant effect on x-rays produced by betatron
oscillations which has been shown to be sensitive to focal
spot asymmetries \cite{mangles09}.

This work has been supported by the BMBF (contract
no.~03ZIK052) and by the DFG (contract no.~TR18). S.P.D.M.
thanks the Royal Society for support and A.G.R.T. was
supported by the NSF under contract 0903557. We also thank
the {\sc Osiris} Consortium.

\end{document}